\documentclass[aps,nofootinbib,showpacs]{revtex4}
\usepackage{amsmath,amssymb,graphicx}

\newcommand{\be}{\begin{eqnarray}}
\newcommand{\ee}{\end{eqnarray}}

\begin{document}
\bibliographystyle{apsrev}

\title{Instanton Contribution to the Proton and Neutron Electric
Form Factors}
\author{P. Faccioli, A. Schwenk and E.V. Shuryak}
\affiliation{Department of Physics and Astronomy, State University
of New York, Stony Brook, NY 11794-3800}
\date{\today}


\begin{abstract}
We study the instanton contribution to the proton and neutron
electric form factors. Using the single instanton approximation,
we perform the calculations in a mixed time-momentum
representation in order to obtain the form factors directly in momentum
space. We find good agreement with the experimentally measured
electric form factor of the proton. For the neutron, our result
falls short of the experimental data. We argue that this
discrepancy is due to the fact that we neglect the contribution of
the sea quarks. We compare to lattice calculations and a
relativistic version of the quark-diquark model.
\end{abstract}


\maketitle

\section{Introduction}
\label{introduction}

Electro-magnetic form factors provide valuable information about
the structure of hadrons and the strong interaction dynamics. At
low momenta, they directly probe the electric and magnetic charge
distribution inside the hadron. In general, the form factors
are related to the elastic amplitude for a given hadron to absorb
a virtual photon. Thus, one can access the interaction responsible for
the recombination of the partons into the hadron.

The electro-magnetic form factors of the nucleon are currently
subject to a renewed experimental interest. At low momenta, the
proton electric and magnetic form factors can be very well
described by the same dipole fit,
$G^p_{E(M)\,dip}=e(\mu)/(1+Q^2/M_{dip}^2)^2$, where
$M_{dip}=0.84~\textrm{GeV}$. For larger momenta ($Q^2 \gtrsim
2~\textrm{GeV}^2$), however, recent measurements at JLab show that
the electric form factor falls off faster than the magnetic
one~\cite{JLAB1,JLAB2}. On the other hand, the electric form
factor of the neutron has been measured up to $Q^2 \approx
2~\textrm{GeV}^2$~\cite{NGE1,NGE2,NGE3,NGE4,NGE5,NGE6,NGE7}. It
was found to be positive, which indicates an inhomogeneous
distribution of the positive and the negative charge in the
neutron, with the positive charge concentrated near the center.

From the theoretical point of view, the nucleon form factor is one
of the few hadronic quantities of fundamental importance. Perturbative
QCD (pQCD) predicts power asymptotics $G \sim 1/Q^4$~\cite{BF} related to
the minimal number (two) of exchanged gluons, which agrees with data. 
However, the coefficient which is related to the so-called light cone 
nucleon wave function (see e.g.~\cite{CZ}) does not agree with the
pQCD asymptotics. This gives rise to the question whether it is actually
described by pQCD at all in the experimentally reached region.

The situation for the charged pion form factor is similar. The 
power of Q matches pQCD, but the coefficient does not, although it is
well known asymptotically. Therefore, it must be
produced by some non-perturbative effect. In our recent paper~\cite{pionFF},
we have calculated the instanton contribution to the pion
form factor using the simplest single instanton approximation
(SIA). This led to a very good description of the data, without any
new parameters involved. In fact, the only relevant parameter is the
average instanton size in the QCD vacuum. We further found that 
the contribution from a single instanton exceeds the one of 
asymptotic pQCD in a rather wide region $Q^2 \lesssim 10~\textrm{GeV}^2$ 
accessible to current experiments. Moreover, the form factor 
of the pion was found to follow closely the well-known vector 
dominance monopole expression.

It is the guiding motivation of the current investigation to see whether the
dipole form of the proton electric form factor can also be
explained using instantons. One possible motivation comes from the
similarity between the pion and the scalar diquark channel, i.e., the
$ud$ content of the nucleon. This similarity becomes exact in
two color QCD and approximate (up to a factor $1/2$) for three colors
in the instanton context (see e.g. the discussion in~\cite{RSSV}). 
More formally, we find that the nucleon form factor also receives 
contributions from maximally enhanced instanton diagrams with two zero 
modes. In contrary, e.g., the form factor of the delta does not
receive similar maximal instanton contributions. In fact, in the SIA
it vanishes. As the main result of this investigation, we indeed find that
the proton electric form factor at intermediate momentum transfer $Q^2
\sim 1-4~\textrm{GeV}^2$ is well reproduced by the single instanton
contribution with the same standard parameters of the instanton ensemble.

There are a number of other interesting questions
which emerge from the experimental information
on the proton form factors. For example, one would like to
understand why the low $Q^2$ data for both $G_E^p(Q^2)$ and
$G_M^p(Q^2)$ follow the same dipole fit. In the case of the pion
form factor, the success of the monopole fit can be understood in
terms of vector meson dominance. However, there is not such a
simple picture which can explain the dipole behavior of the proton
form factors, although the mass in the dipole fit, $M_{dip}$, is
close to the vector meson mass.

There are also interesting theoretical questions arising form the
data for the neutron form factor. The fact that the electric form
factor is non-zero is a clean signature that a naive
non-relativistic quark model description based on $SU(6)$
symmetry is inadequate for the dynamic properties of the nucleon.
It is then remarkable that such a picture works so well in
describing the neutron to proton magnetic moment
ratio. Clearly, in order to solve this puzzle, we need to
understand what is the main source of $SU(6)$ breaking in the
neutron wave function.

Such questions have been addressed in a number of phenomenological
models as well as in lattice simulations~\cite{martinelli,draper}.
In particular, we discuss two recent works which are related to
our analysis. Dong \emph{et al.} calculated the proton and
neutron form factors in lattice QCD and studied the sea quark
contribution~\cite{Lattice1} which is due to
disconnected diagrams. It was found that the proton electric
form factor can be very well reproduced by the connected
components of the relevant Green functions, with the contribution
of the sea being negligible. On the other hand, the connected
contribution accounts for roughly only half of the neutron
electric form factor. Therefore, their analysis shows that the sea
cannot account for the entire $SU(6)$ breaking observed in the
neutron wave function.

Ma \emph{et al.} considered a different source of $SU(6)$
breaking~\cite{LFdiquark}. They calculated the form factors of the nucleon
in a simple quark-diquark model, in which relativistic covariance
is enforced by using the light front dynamics formalism.
In this model, they considered trial wave functions depending on
phenomenological parameters, which were fixed in order to
reproduce the static properties of the nucleon. By varying such
parameters, they could tune the amount of $SU(6)$ breaking in their
valence picture. They found that such a simple model can reproduce very
well the existing data for the form factors of the nucleon up to
$Q^2 \sim 2~\textrm{GeV}^2$. Moreover, in this model, the ratio
$G_E^p/G_M^p$ at high momenta is a consequence of the relativistic
correlation between the spin and the momentum of the constituents.

We now come back and discuss the significance of the instanton
contribution to the neutron and proton
electric form factors. Instantons are topologically non-trivial
solutions of the Euclidean Yang-Mills equation of motion.
Physically, they describe tunneling events in the QCD vacuum and
are associated with strong non-perturbative color fields. Form
factors are amplitudes of a hadron retaining its identity after
large momentum transfers and the strong fields of the instantons may
transfer momentum between several quarks at once. Furthermore,
instanton zero modes lead to a special importance of instantons for
any problems involving light fermions, especially in spin-zero channels
such as pions or diquarks.

In the Instanton Liquid Model (ILM) (for a review
see~\cite{shuryakrev}), the QCD partition function is assumed to
be saturated by an ensemble of instantons and anti-instantons of a
typical size, $\rho \approx 1/3~\textrm{fm}$, and a typical
density, $n \approx 1~\textrm{fm}^{-4}$. Previous works have shown
that the ILM describes quantitatively the spectrum of the light
mesons~\cite{corrmesons} and baryons~\cite{corrbaryons}. The instanton
induced interaction is effective only between quarks of different
flavor and chirality. This implies that, in the nucleon, only two
quarks can be bound by the 't~Hooft interaction, while the third
one is more loosely bound.
Indeed, it was shown~\cite{corrbaryons} that the $u$ and $d$ quark in the
nucleon form a bound state, a scalar diquark with a mass comparable to
that of a constituent quark. This analysis provided a microscopic
motivation for the quark-diquark model of the nucleon.

There are, however, some differences between the quark-diquark
picture which emerges from the ILM and the simplest phenomenological
model described above. For example, Ma \emph{et al.} considered a
nucleon wave function with an equal mixture of a scalar and a
vector diquark~\cite{LFdiquark}, while the 't~Hooft interaction
generates a scalar diquark only. Moreover, in~\cite{LFdiquark},
the diquark is treated as a point-like particle, while in the ILM
its size has been estimated to be approximately
$0.4~\textrm{fm}$~\cite{3ptILM}. Finally, in the ILM it is possible
to account for the contribution of the meson cloud and
the sea quarks simultaneously, both contributing to the $SU(6)$
symmetry breaking of the nucleon wave function.

In a previous work, two of the authors calculated the proton
electro-magnetic three-point function in coordinate
space~\cite{3ptILM} both from numerical simulations in the ILM, i.e.,
including multi-instanton effects, and
analytically in the SIA. The Green function evaluated
theoretically was then compared to a phenomenological one derived 
from the Fourier transform of several
parametrizations of the experimental data. This approach had the
advantage to consider large-sized correlation functions, for which
the contribution of the continuum of excitations was certainly
negligible. However, such a procedure has the shortcoming that it
does not allow a direct comparison to the experimental data. From
a theoretical point of view, the main result was that the proton
electro-magnetic three-point function is completely dominated by
the contribution of a single instanton, up to surprisingly large
distances of $\approx 1.8~\textrm{fm}$. From a phenomenological
point of view, it was shown that the ILM predictions are consistent
with a deviation of $G_E^p$ from the dipole fit. This is in nice
agreement with the result obtained by Ma \emph{et al.} in their
simple model.

In the present work we develop a much simpler single instanton
calculation, based
on the time-momentum correlators used in~\cite{pionFF}
and~\cite{mymasses}. Such a scheme accounts for the leading
single instanton effects. The calculation is physically very
transparent and presents several analogies with perturbation
theory. Moreover, we compare directly to the experimental data.
In the pion and nucleon channel, the single instanton contribution
is certainly dominant for hadronic processes at intermediate
momenta, $Q^2 \geq 1 \, \text{\textrm{GeV}}^{2}$. Moreover, it
constitutes the relevant gluonic configurations, which take over
with the breakdown of perturbation theory.

We have not calculated the contribution of the quark sea (the
disconnected diagrams) to the form factor, but argue it to be
small, roughly $\sim 1/10$ of the proton form factor and
half of the neutron form factor. We will
proceed by describing the details of our calculation. In
section~\ref{discussion}, we show our results and close by
discussing the physical implications.

\section{Calculational Setup}
\label{calculation}

For the determination of the electric form
factors of the nucleon, we follow the method used in the
calculation of the charged pion electro-magnetic form
factor~\cite{pionFF}. In the wall-to-wall (W2W) formalism, the
electric form factors can be extracted from a combination of
three- to two-point functions. In particular, we choose to work in
the Breit frame and consider the following spatial Fourier
transform of the Euclidean three-point correlator
\be
\label{G3}
G_4(t,{\bf q}/2;-t,-{\bf q}/2) =
\int d^3{\bf x} \, d^3{\bf y} \, e^{i \, {\bf q} \cdot ({\bf x} +
{\bf y})/2} \, \langle 0 | \, \text{Tr} \,
\eta_{\text{sc}}(t,{\bf y}) \, J_4(0,{\bf 0}) \,
\bar{\eta}_{\text{sc}}(-t,{\bf x}) \, \gamma_4 \, | 0 \rangle .
\ee
$J_4$ is the fourth component of the electro-magnetic current
operator and $\eta_{\text{s}}(x)$ is the so-called nucleon scalar
current which, in the case of the proton, reads~\footnote{We note
that the nucleon scalar current contains explicitly the operator which
excites a scalar $ud$ diquark. It is this combination that couples strongly
to the instanton zero modes. The corresponding operator for the
neutron is obtained through the substitution $u \leftrightarrow d$.}
\be
\label{current}
\eta_{\text{s}}(x) = \epsilon_{abc} \, [ u^a(x) C \gamma_5 d^b(x) ] \,
u^c(x) .
\ee
Accordingly, we evaluate the Fourier transform of the nucleon
two-point given by
\be
\label{G2}
G(t,{\bf q}) = \int d^3{\bf x}
\, e^{i \, {\bf q} \cdot {\bf x}} \, \langle 0 | \, \text{Tr} \,
\eta_{\text{sc}}(t,{\bf x}) \, \bar{\eta}_{\text{sc}}(0) \,
\gamma_4 \, | 0 \rangle .
\ee
In both Eq.~(\ref{G3}) and~(\ref{G2}), the additional $\gamma_4$ has
been inserted in order for the correlators to receive maximal
single-instanton contribution (see the discussion in~\cite{3ptILM}
and below).

For large Euclidean times, one can isolate the contribution of the
lowest lying state to the Green function
\be
\label{spectralG3}
G^{p(n)}_4(t,{\bf q}/2;-t,-{\bf q}/2) \rightarrow 2
\, \Lambda^2_{\text{sc}} \, \biggl( \frac{M}{\omega_{{\bf q}/2}}
\biggr)^2 \, G^{p(n)}_{\text{E}}(Q^2) \, e^{-2 \,
\omega_{{\bf q}/2} \, t} ,
\ee
where $G^{p(n)}_{\text{E}}(Q^2)$ denotes the proton (neutron)
electric form factor and $\Lambda_{sc}$ the coupling of the scalar
current, Eq.~(\ref{current}), to the nucleon. The nucleon pole
in the two-point function is similarly reached
\be
\label{spectralG2}
G(t,{\bf q}) \longrightarrow 2 \,
\Lambda^2_{\text{sc}} \, e^{-\omega_{\bf q} \, t} .
\ee

In order to calculate the instanton contribution to such
correlation functions, we use the SIA, which was introduced
in~\cite{shuryak82} and studied in detail in~\cite{SIA}. In such an
approach, only the contribution from the closest instanton is
taken explicitly into account, while the effects of the
other instantons are incorporated in two induced parameters, the quark
effective mass, $m^\star \approx 85~\textrm{MeV}$, and the average
instanton density, $\bar{n} \approx 1~\textrm{fm}^{-4}$.

The quark propagator in the instanton background is known
exactly~\cite{SIprop} and consists of a zero-mode part and a
non-zero mode part, $S^I(x,y)=S^I_{zm}(x,y)+S^I_{nzm}(x,y)$. We
showed that the SIA is reliable only if the relevant Green functions
receive contribution from more than one zero-mode
propagator~\cite{SIA}. In fact, the additional $\gamma_4$ matrix in
Eq.~(\ref{G3}) and~(\ref{G2}) has been inserted in order to guarantee
such an enhancement.

\begin{figure}
\includegraphics[scale=0.7,clip=]{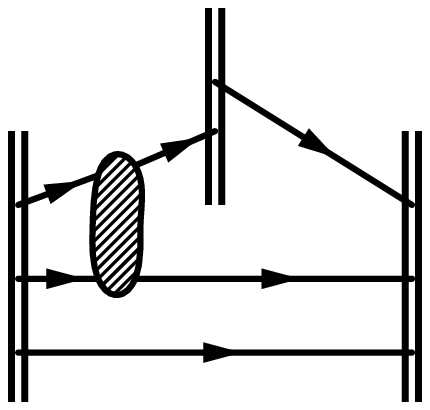}
\hspace*{1.5cm}
\includegraphics[scale=0.7,clip=]{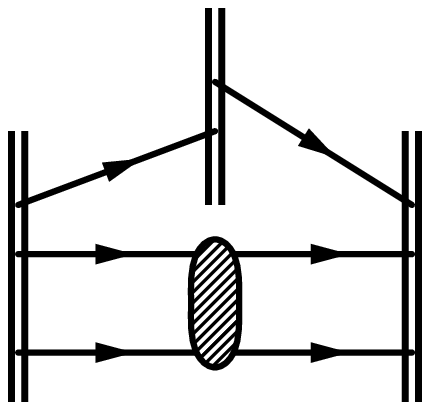}
\hspace*{1.5cm}
\includegraphics[scale=0.7,clip=]{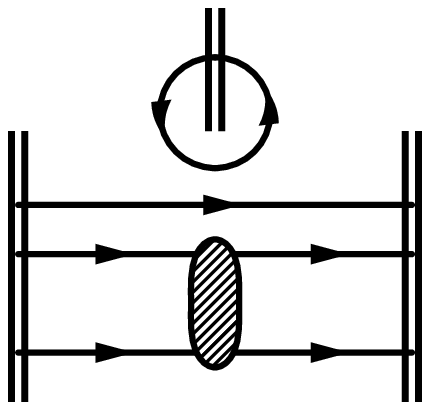} \\
(A) \hspace*{4.6cm} (B) \hspace*{4.6cm} (C) \\
\caption{Graphical representation of the typical contributions
to the W2W nucleon electro-magnetic three point
function. The double lined ``walls'' correspond to the
spatial Fourier integration, see~\cite{pionFF}.
The dashed ellipse denotes the four quark (zero-mode) instanton
interaction. The nucleon is excited at the left, struck by the
virtual photon in the middle and absorbed at the right. Two
contributions to the connected three-point function are shown.
Diagram~(A) probes the diquark content of the nucleon, whereas in
diagram~(B), the photon interacts with the remaining
quark. Diagram~(C) is disconnected, where the photon probes the sea quark
content of the nucleon.}
\label{wwfig}
\end{figure}

In this work, we choose to further simplify the calculation by
adapting the so called "zero-mode approximation", in which the
non-zero mode part of the propagator is replaced by the free one,
$S^I(x,y) \simeq S^I_{zm}(x,y)+S_0(x,y)$. This approximation is
accurate in the case of the nucleon three- and two-point functions
which we are considering~\cite{3ptILM}. Finally, it is convenient to
work directly in a time-momentum representation for the Green
functions. This is achieved by using the W2W quark propagators
evaluated in~\cite{pionFF}. Two typical diagrams contributing
to the connected three-point function, Eq.~(\ref{G3}), in the SIA are shown in
Fig.~\ref{wwfig}~(A) and~(B). Due to the chiral and flavor structure of the
instanton induced interaction, the zero modes are restricted to the
$u$ and $d$ quark inside the nucleon. This reduces the possible
diagrams to the structure depicted in Fig.~\ref{wwfig}~(A) and~(B) and
a similar diagram with the instanton to the right. The total
contribution is nevertheless still quite involved due to the
charge conjugation matrix in the nucleon currents. The interpretation
is that in diagram~(A) of Fig.~\ref{wwfig} the virtual photon probes
the diquark content of the nucleon, whereas in diagram~(B) the photon
interacts with the residually bound quark.

After Wick contraction, the calculation of the three- and
two-point functions, Eq.~(\ref{G3}) and~(\ref{G2}), reduces to
evaluating the averages of traces of W2W propagators in the single
instanton (and anti-instanton) background. The average involves an
integration over the position, size and color orientation of the
instanton. We approximate the instanton size distribution to be
simply $d(\rho) =\bar{n}~\delta(\rho-\bar{\rho})$, with a typical
instanton density, $\bar{n} = 1~\textrm{fm}^{-4}$, and a typical
instanton size, $\bar{\rho} = 1/3~\textrm{fm}$ taken from the 
ILM.\footnote{In~\cite{pionFF}, we evaluated the pion form factor
using different parametrizations of the instanton size
distribution. We found that the predictions obtained from such a
simple delta-function ansatz agree within $10 \%$ 
with those obtained using a parametrization of the lattice 
results for $d(\rho)$.} In general, Wick contracting generates 
connected as well as disconnected averages.

The disconnected contribution, Fig.~\ref{wwfig}~(C), requires 
some discussion. Physically, it corresponds to the 
effects of the sea quarks on the form factor. In perturbation theory,
this contribution is subleading at large $Q$ due to
the additional gluon exchange required to transfer momentum
from the struck sea quark to the valence one. In the instanton
background field, this momentum transfer can occur via the 
instanton field itself~\footnote{The chiral structure of the instanton
interaction requires a non-zero mode propagator for the sea quark 
loop. The free propagator used in the zero-mode 
approximation would lead to a vanishing
contribution.} and also leads to an extra suppression at
large $Q$. The total momentum flowing into the sea quark loop
${\bf q}$ has to be transferred to the valence quarks, which leads
to an extra ``instanton form factor'' $\exp(-\rho \, |{\bf q}|)$
from the non-zero mode propagator in the sea quark loop. This small
factor appears in addition to the form factors from the zero mode 
and non-zero mode propagators in the valence
pieces.\footnote{These go as $\exp(-\rho \, |{\bf q}|/6)$ since
every valence quark has a total momentum $|{\bf q}|/6$ on average
in the Breit frame.}

Unfortunately, the evaluation of the W2W non-zero mode propagator 
is extremely involved and we refrain from calculating the 
contribution of the disconnected diagram, Fig.~\ref{wwfig}~(C). 
We will argue below that it is indeed a small contribution to the
proton form factor.

In this work, we focus on the {\it electric} form factors of the proton and
the neutron only, as they come from maximally enhanced diagrams. The
instanton contribution to the {\it magnetic} form factors of the nucleon can be
extracted from a different combination of the three- and two-point
functions~\cite{draper,Lattice1}, which however
receives only subleading contributions from a single instanton.

Once the Green functions are evaluated, the electric form factor
can in principle be determined from the ratio of the three- to
two-point correlators in the large Euclidean time limit, $t \to \infty$.
However, in the approach of this work focused on a single instanton,
one cannot choose the time interval to be
arbitrarily large. In fact, as $t$ becomes comparable with
the typical distance between instantons, $\bar{n}^{-1/4} =
1~\textrm{fm}$, many-instanton effects should begin to play a
non-negligible role and the SIA will eventually break down. The
SIA calculation of the three- and two-point functions are
compared to numerical simulations in the ILM in~\cite{3ptILM}.
We verified that the SIA calculation of the two-point
correlator, Eq.~(\ref{G2}), is reliable up to distances of
$\approx 0.9~\textrm{fm}$. On the other hand, we found that the
single instanton contribution to the three-point function,
Eq.~(\ref{G3}), saturates the ILM results up to much larger 
distances of $\approx 1.8~\textrm{fm}$.

In order to access the proton and neutron form factors, we need to ensure
that, at the largest value of the Euclidean time allowed by the
SIA, the nucleon pole is reasonably isolated in both the two- and
three-point correlation functions. This was assured for the pion,
essentially due to the large separation from its resonances. For
the nucleon, this is a more delicate task, because the first
resonance with the same quantum numbers, $N(1440)$, is only a few
hundred MeV heavier than the ground state. This implies that
larger time intervals are needed to separate their contributions. 
Therefore, it is a priori not
guaranteed that there is a window, in which the SIA is reliable and
the nucleon is isolated from its resonances.

\begin{figure}
\includegraphics[scale=0.8,clip=]{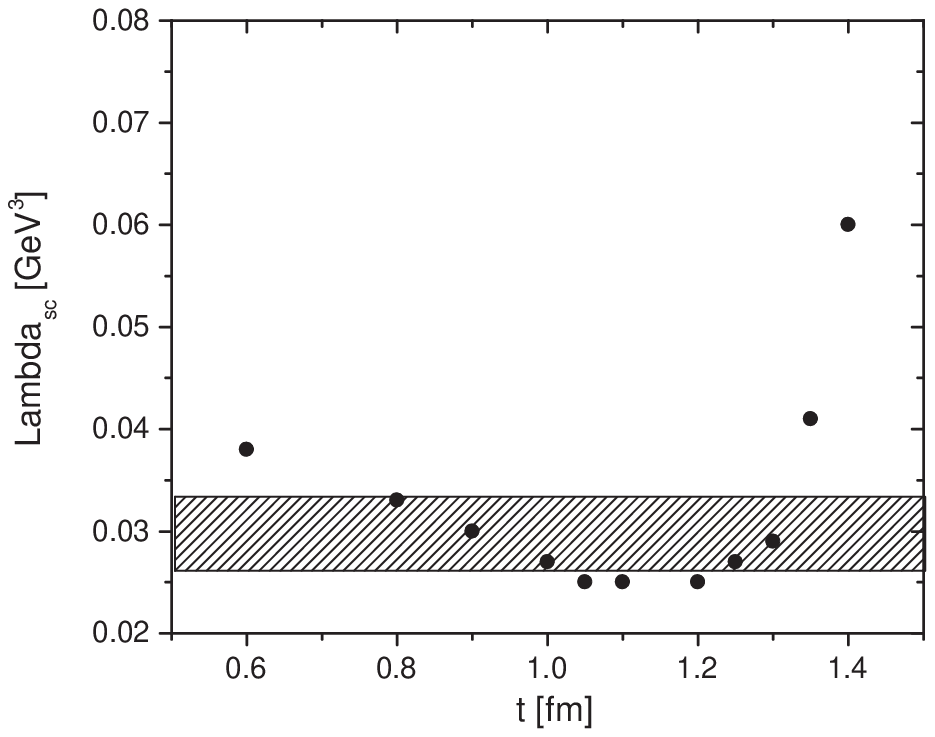}
\includegraphics[scale=0.8,clip=]{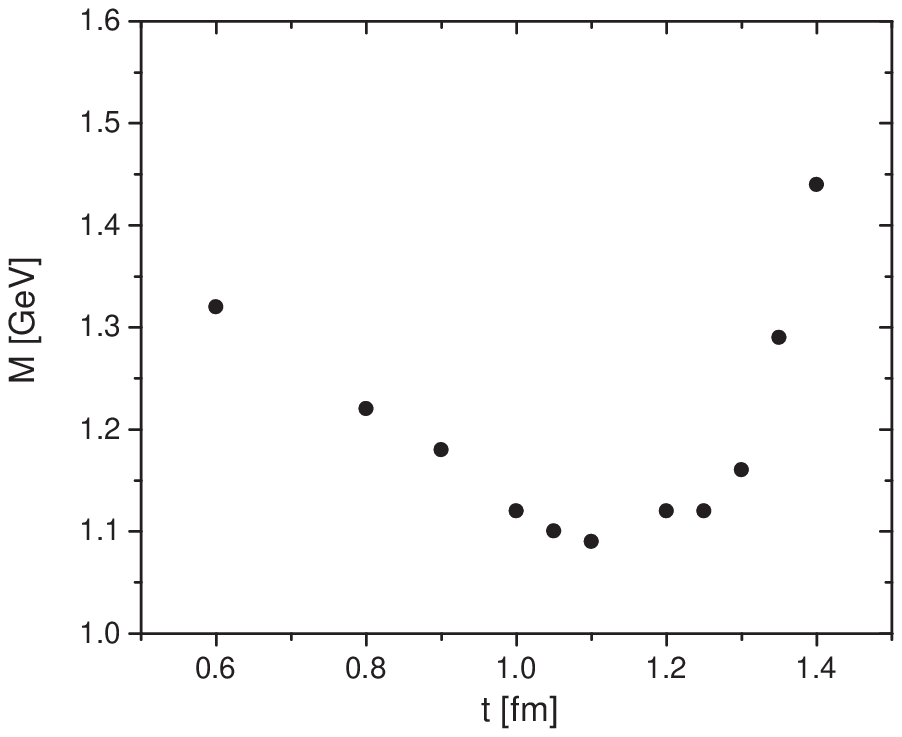}
\caption{The scalar coupling to the nucleon,
$\Lambda_{\text{sc}}$, (left) and the nucleon mass (right)
evaluated from the SIA two-point function for different Euclidean
times. The shaded area in the left figure represents the range of
values of $\Lambda_{\text{sc}}$ obtained from quenched and
unquenched lattice simulations, e.g.~\cite{Lattice2}.}
\label{parametersvsT}
\end{figure}

In Fig.~\ref{parametersvsT}, we show the SIA results for the
nucleon coupling and mass, obtained from a fit of the
two-point function, Eq.~(\ref{G2}), retaining only the ground
state in the spectral decomposition, as in Eq.~(\ref{spectralG2}).
We find that a plateau, which indicates the complete isolation of
the proton signal, is obtained for $t \gtrsim 1~\textrm{fm}$.
Moreover, as $t$ becomes larger than $1.3~\textrm{fm}$, the single
instanton contribution rapidly dies out and the approximation
breaks down. Unfortunately, on the basis of the analysis developed
in~\cite{3ptILM}, we estimate the maximal time interval for which
we can trust our two-point function SIA calculation to be $t \approx
0.9~\textrm{fm}$.\footnote{We note that the total time interval in
the three-point function is given by $2 \, t$.} This implies that,
in our calculation, the nucleon is not completely isolated and we
expect a small contamination of the correlation functions from
excited states. From the two plots in Fig.~\ref{parametersvsT}, we
estimate that such effects result in corrections of the order of
$10 \%$.

We extract the electric form factors from the ratio
\be
\label{GE}
2 \, \Lambda_{\text{sc}}^2(t) \, \biggl( \frac{\omega_{{\bf
q}/2}(t)}{M(t)} \biggr)^2 \, \frac{G^{p(n)}_4(t,{\bf
q}/2,-t,-{\bf{q/2},t})}{(G(t,{\bf q}/2))^2} \to G^{p(n)}_E(Q^2) ,
\ee
where $\Lambda_{\text{sc}}(t)$, $M(t)$ and $\omega_{{\bf
q}/2}(t)=\sqrt{{\bf q}^2/4+M(t)^2}$ denote the values extracted from
a fit of the two point function $G(t,{\bf q}/2)$ keeping only the
nucleon contribution in the spectral decomposition. The ratio of
$G_4(2t)/G(t)^2$ ensures that both correlators can be calculated
reliably in the SIA. The two factors of the two-point function are
needed to sustain the nucleon pole over the total Euclidean distance
$2 \, t$ with a necessary $t = 0.9~\textrm{fm}$. It corresponds to the
leading order in a virial expansion in the instanton density. This
procedure is at the expense of a dependence on the multi-instanton
induced parameters, namely the quark effective mass $m^\star$ and the
average instanton density $\bar{n}$, or equivalently a dependence on
the nucleon coupling $\Lambda_{\text{sc}}$ and mass. For the pion form
factor, it could be achieved that such a dependence cancels in the
calculation, with only the average instanton size $\bar{\rho}$ remaining.

\begin{figure}
\includegraphics[scale=0.8,clip=]{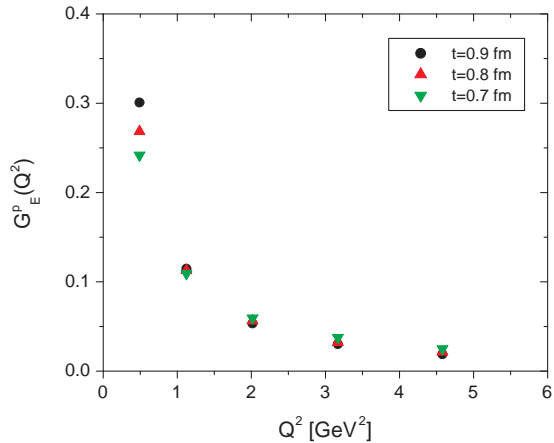}
\caption{The proton electric form factor $G_E^p$ as a function of
Euclidean time $t$. Although the relevant Green functions
receive a contribution from the excited states of the order of $10\%$,
for $t=0.9~\textrm{fm}$, the form factor is independent of $t$.}
\label{GEvsT}
\end{figure}

In Fig.~\ref{GEvsT}, we give the resulting proton electric form factor
in the SIA, obtained for different values of the Euclidean time.
We observe that our outcome is nearly independent on the chosen time
interval restricted to the SIA window. This indicates a cancellation
between the small contribution from the excited states to the
numerator and the denominator in Eq.~(\ref{GE}).

\section{Results and Discussion}
\label{discussion}

\begin{figure}
\includegraphics[scale=0.8,clip=]{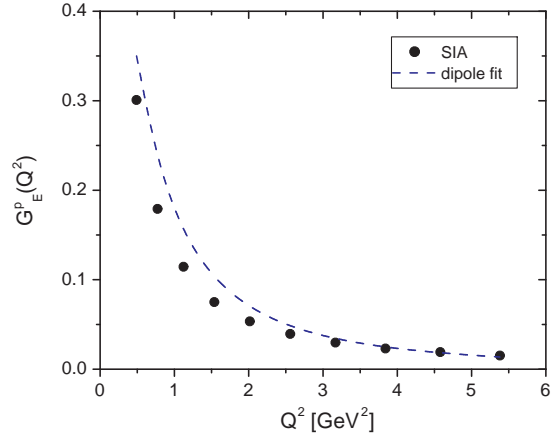}
\caption{The proton electric form factor evaluated in the SIA
(points) compared to the dipole fit (dashed line). The shown
results are obtained using $t = 0.9~\textrm{fm}$. The SIA results
are reliable for $1~\textrm{GeV}^2 \lesssim Q^2 \lesssim 4~\textrm{GeV}^2$
(see the discussion in the text and in~\cite{mymasses,pionFF}).}
\label{ProtonFF}
\end{figure}

The aim of the present calculation is to show that the dipole behavior
of the proton electric form factor can be explained from the
interaction of the quarks with the field of a single 
instanton. Our final result for $G^p_E$ is presented in
Fig.~\ref{ProtonFF} in comparison with the dipole
fit.\footnote{At large momenta, the electric form factor of the
proton is experimentally less well resolved than both its ratio to
the magnetic one and the magnetic form factor itself. From the
measurements at JLab~\cite{JLAB1,JLAB2}, one finds that at $Q^2
\sim 1-4~\textrm{GeV}^2$ the electric form factor is reduced by about
$20-50\%$ relative to the magnetic one, whereas the magnetic form
factor is still reproduced by the dipole fit very well.} The
results show that the instanton-induced contribution can account
for the correct magnitude of $G_E^p$. Thus, it should be included
in any dynamical model. Whether there is indeed a direct relation
between the nucleon form factor and instantons in the QCD vacuum can be
further tested on the lattice, in many ways and on a
configuration-per-configuration basis.  
We do not wish to claim that the SIA results have the
accuracy to explain the high $Q^2$ precision data for the ratio of
$G_E^p/G_M^p$. However, much more involved numerical simulations have
shown that the ILM is consistent with the high momentum deviation of
the proton electric form factor from the dipole fit~\cite{3ptILM}.

\begin{figure}
\includegraphics[scale=0.8,clip=]{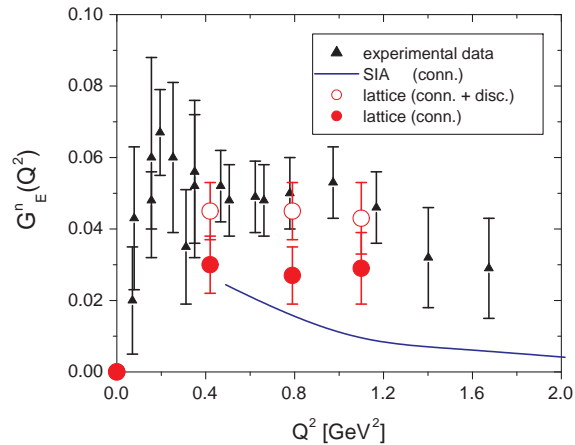}
\caption{The neutron electric form factor in the SIA (solid line
again for $t = 0.9~\textrm{fm}$) in comparison to the experimental
data~\cite{NGE1,NGE2,NGE3,NGE4,NGE5,NGE6,NGE7} (triangles),
lattice results retaining only the connected components (filled
circles) as well as both connected and disconnected parts (empty
circles)~\cite{Lattice1}. The SIA results are reliable for
$Q^2 \gtrsim 1~\textrm{GeV}^2$ (see the discussion in the text and
in~\cite{mymasses,pionFF}).}
\label{NeutronFF}
\end{figure}

The result for the neutron form factor is presented in
Fig.~\ref{NeutronFF} in comparison to the available experimental
data and the lattice results of Dong \emph{et
al.}~\cite{Lattice1}. As expected, the presence of a massive
diquark generates an inhomogeneous charge distribution and leads
to a positive form factor. However, in the instanton model, the masses
of the scalar diquark and the constituent quark are not that different,
which leads to a small neutron form factor. Furthermore, we
observe that the connected component of the Green function
accounts for roughly half of the form factor only. This is in
qualitative agreement with lattice results, which also found
that half of the neutron form factor comes from the disconnected
diagrams we have neglected in the present calculation. This
provides an estimate for its value and supports our assumptions
that it is small compared to the proton form factor. In the
instanton model it is subleading as explained above. 

Summarizing, we have computed the single instanton contribution to
nucleon electric form factors. We have carried out the calculations
analytically in a window of momentum transfer in which the
single instanton approximation can be used. The range of validity of
the SIA for the nucleon was found to correspond to momentum 
transfers $Q^2 \sim 1-4~\textrm{GeV}^2$. At smaller $Q^2$, 
multi-instanton effects appear, and at larger $Q^2$, there are 
admixtures of excited states (see the discussion in~\cite{mymasses}). 
Already the existence of such a window for the nucleon in the SIA 
is highly non-trivial due to the presence of a rich and close 
spectrum of excited states. The only other hadron for which this was
shown to be the case is the pion~\cite{pionFF}. 
The physical reason is that the (single) instanton 
induces a compact quark-diquark structure of the 
nucleon. Other baryons, such as the decouplet members, do not have
such a structure and thus are expected to have smaller form
factors. For numerical results we used the standard parameters of 
the instanton ensemble~\cite{shuryak82}, 
fixed 20 years ago. A similar analysis of the proton to delta
transition form factors is in preparation~\cite{Ndelta}.

\end{document}